\documentclass[12pt]{article}
\usepackage{amsmath}
\usepackage{amssymb}

\usepackage{subfigure}
\usepackage{psfrag}
\usepackage{epsfig}
\usepackage{graphicx}
\usepackage{tikz}
\usepackage{cite}
\usepackage{hyperref}

\newcount\xrefpos \xrefpos=0
\newcount\yrefpos \yrefpos=0
\newcount\xput \xput=0
\newcount\yput \yput=0

\def\refpos#1 #2 #3{\global\xrefpos=#1 \global\yrefpos=#2
                         \rlap{$\smash{#3}$}}
\def\put #1 #2 #3{\xput=#1 \yput=#2
                  \advance\xput by -\xrefpos
                  \advance\yput by -\yrefpos
                  \rlap{\kern\the\xput truebp
                        \vbox to 0pt{\vss\hbox{$\displaystyle #3$}
                        \kern\the\yput truebp}}}
\def\beginlabels\refpos#1\endlabels{\hbox{$\refpos#1$}}

\newcommand{\ba}{\begin{eqnarray}}
\newcommand{\ea}{\end{eqnarray}}
\newcommand{\beq}{\begin{equation}}
\newcommand{\eeq}{\end{equation}}

\usepackage{amsfonts,amssymb}

\begin{document}

 \begin{center}
 {\Large \bf Competing Holographic Orders}

\bigskip
\bigskip
\bigskip
\bigskip

\vspace{3mm}

Pallab Basu$^{a,}$\footnote{email: pallab@phas.ubc.ca} , Jianyang He $^{a,}$\footnote{email: jyhe@phas.ubc.ca}, Anindya Mukherjee$^{a,}$\footnote{email: anindya@phas.ubc.ca}, Moshe Rozali $^{a,}$\footnote{email: rozali@phas.ubc.ca} and Hsien-Hang Shieh$^{a,b.}$\footnote{email:bshieh@physics.ucla.edu}

\bigskip \centerline{$^a$\it Department of Physics and Astronomy}
\smallskip\centerline{\it University of British Columbia}
\smallskip\centerline{\it 6224 Agricultural Road, Vancouver, B.C.}
\smallskip\centerline{\it Canada  V6T 1Z1}
\bigskip\medskip
\centerline{$^b$\it Department of Physics and Astronomy}
\smallskip\centerline{\it University of Cliafornia, Los Angeles}
\smallskip\centerline{\it CA, 90095, USA}

 \end{center}

 \bigskip\bigskip\bigskip


\abstract{We model competition between different macroscopic orders in an holographic context. The orders we considered are a superconducting order, modeled by a charged scalar field, and a magnetic order modeled by a neutral scalar field. We also discuss the case of two competing scalars coupled to a single gauge field.

In all cases discussed here the phases tend to compete, rather than enhance each other. The condensation of one scalar hinders any further instabilities, unless we have a sufficiently strong repulsive interactions between the bulk scalars. We provide both analytic arguments and numerical demonstration of this fact.

Based on the cases discussed here, we conjecture that holographic orders tend to compete for attractive bulk interactions, including gravity, and to cooperate, or be mutually enhancing, for repulsive bulk interactions between the corresponding order parameters. }
%

\newpage
\section{Introduction and Conclusions}

Holographic Condensed Matter (for a review see e.g \cite{Sachdev:2010p11089}) is a new approach to modelling strongly interacting systems and calculating their properties. The method involves a set of tools acquired through investigations of quantum gravity, including string theory, black hole physics and higher dimensional gravitational theories. As such, better understanding of the scope and usefulness of the approach for various physical systems may improve our understanding of quantum gravity. However, the main utility of this approach for the time being is the construction of new classes of strongly interacting universality classes, which may well be useful to understanding and modeling real materials, in cases where conventional calculational tools fail.

With this phenomenological approach in mind, we discuss in this paper the issue of competing orders.  The holographic approach, by and large, has concentrated so far on the dynamics of a single order parameter  coupled to all relevant conserved quantities (somewhat similar to the Landau-Lifshitz theory of phase transitions, however not necessarily near the phase transition point). Strongly correlated electron systems tend to have complicated phase diagrams, with many possible orders, such as magnetic orders, striped phases or superconductivity. This is in contrast to conventional Fermi liquids, which tends to have very few instabilities.  It is interesting therefore to look at the interplay of various orders in the holographic context, to model specific phase diagrams, and even perhaps to draw some general conclusions.

In  weakly interacting Fermi liquids, the onset of one type of order tends to produce a mass gap (at least for parts of the Fermi surface), thus inhibiting further instabilities. In this sense the phases tend to compete, the parameter range of a  potential instability tends to shrink when the competing order sets in, for example the critical temperature is lowered. In some cases, the onset of one type of order prevents any further instability. This is one reason for the robustness and usefulness of the Fermi liquid picture.

On the other hand, quantum critical points may change that picture. Such phase transitions deviate from the Landau paradigm of phase transitions in that the critical theory may have new light degrees of freedom, beyond the order parameter and the conserved quantities. These emergent degrees of freedom (for example the emergent gauge fields in \cite{Senthil:2004p10637}) can enhance some instabilities. For example, it is speculated that new massless modes at a quantum critical point may replace the phonons of the BCS theory as the pairing mechanism for superconductivity.

In this note we initiate a discussion of competing orders in the holographic context, and examine a few models. By and large we find that the conventional picture is still valid for holographic models. That is, for generic models the onset of one order tends to suppress all other potential orders. We do find some exceptions to this rule, and comment on the set of conditions  that seem to be needed to realize holographically the compelling idea of mutually enhancing orders.

The set of quantum phase transitions discussed here is subject to  the general considerations of \cite{Kaplan:2009kr}, in that they are triggered by a bulk field falling below the BF bound, and thus are expected to exhibit BKT-like scaling near the phase transition point. Some holographic models, with this type of mechanism for instability, indeed show the expected scaling behavior \cite{Jensen:2010ga, Iqbal:2010p11098, Jensen:2010vx}. It would be interesting to show such behavior in the present set of models as well.

The plan of the paper is as follows: we present the set of models we use in this paper in section 2, including a brief review of the realization of superconductivity and magnetic order in the holographic context. Section 3 is devoted to the interplay of the superconducting and magnetic orders, as realized in the holographic context. Finally, in section 4 we investigate similar issues in a model of a single order (global U(1) symmetry) coupled to two competing scalar fields.

\section{Background and Setup}

\subsection{Holographic Superconductors at Zero Temperature}

Holographic superconductors are gravity backgrounds coupled to a gauge field (representing global or weakly gauged U(1) in the dual theory), and a charged scalar field whose condensation triggers superconductivity. The specific set of models we consider here was constructed by Gubser and Nellore {\cite{Gubser:2009p11021}} and also by Horowitz and Roberts {\cite{Horowitz:2009p10812}}. These models correspond to a 2+1 dimensional CFTs, at finite density and zero temperature. The presence of chemical potential breaks the conformal symmetry, and the theory flows to an IR theory with Lifshitz scaling {\cite{Kachru:2008p10773}}, with dynamical exponent $z$. The limiting case of $z=1$ corresponds to relativistic CFT in the IR, whereas the case of $z=\infty$ is similar to the one discussed in {\cite{Iqbal:2010p11098}} (which discussed the charged black hole without scalar fields), in the sense of having an $AdS_{2}$ space in the IR.

Consider then the bulk theory described by gravity coupled to a Maxwell field with the Lagrangian:
\begin{equation}
\label{eq:lagrangian}
\mathcal{L} = \frac{1}{2\kappa^{2}}\left[R + \frac{6}{L^2} - \frac{1}{4}\,F^{\mu\nu}F_{\mu\nu}  - |D_{\mu}\psi|^{2}- V(\psi,\psi^{*})  \right],
\end{equation}
where $F_{{\mu\nu}}$ is the electromagnetic field strength, and $A_{0}= \Phi(r)$ is the electric potential (satisfying $A(\infty)=\mu$ at finite chemical potential $\mu$, note that at $T=0$ all non-zero chemical potentials are related by scaling). The charged scalar $\psi$ of charge $q$ triggers superconductivity when condensing. When discussing this set of models, we work exclusively with the quadratic potential $V(\psi,\psi^{*}) = m_{\psi}^{2} |\psi|^{2}$ with $m_{\psi}^{2} > 0$.

The Lagrangian allows for solutions with Lifshitz scaling, with the metric
\begin{equation} ds^{2}= -g(r)^{2} dt^{2}+ \frac{r^{2}}{L_{0}^{2}} (dx^{2}+dy^{2}) + e^{2B(r)} \frac{L_{0}^{2}}{r^{2}} dr^{2} \end{equation}

with
\begin{eqnarray}
g(r)&=& (\frac{r}{L_{0}})^{z} \nonumber\\
B(r)&=&0 \nonumber\\
\psi(r) &=& \psi_{0} \nonumber\\
\Phi(r) &=& \sqrt{2-\frac{2}{z}} \,(\frac{r}{L_{0}})^{z}
\end{eqnarray}

where the parameters of the solution are given by
\begin{equation}
q^{2}= \frac{z m_{\psi}^{2}}{2(z-1)}~~~~\psi_{0} =\frac{2 \sqrt{3}}{m_{\psi}L}\sqrt{\frac{z-1}{(z+1)(z+2)}}~~~~L_{0}= L \sqrt{\frac{(z+1)(z+2)}{6}}
\end{equation}

Following {\cite{Gubser:2009p11021}} we parametrize the solutions by a pair $(z>1, m_{\psi}^{2}>0)$. There exists a solution for each such pair, though such solution is not necessarily obtainable as the IR limit of a UV CFT at finite density. Under conditions discussed in detail in {\cite{Gubser:2009p11021}}, there are irrelevant perturbations that can be potentially used to connect the solutions for appropriate $(z,m^{2})$ to a complete flow from a relativistic 2+1 dimensional CFT.  We have obtained numerically such flows for the cases discussed below.

For later use, let us consider the issue of the Breitenlohner-Freedman (BF) bound \cite{Breitenlohner:1982bm, Breitenlohner:1982jf} in this set of backgrounds. A neutral scalar field $\xi$ satisfies the equation of motion \begin{equation}r^2 \xi ''(r) +(3+z) r \xi '(r) - L_{0}^{2}\,U'(\xi(r))=0 \end{equation} Therefore the scaling dimension $\Delta_{{IR}}$ for small fluctuations of the field $\xi$ satisfies the equation \begin{equation} \Delta (\Delta-1) + (3+z) \Delta - L_{0}^{2} \,m_{{eff}}^{2}=0 \end{equation} where $m_{eff}^{2}$ is the curvature of the potential $U(\xi)$ around the background value of $\xi$. To avoid complex scaling dimensions, which would signal instability, we have the BF bound \begin{equation}  m_{eff}^{2} \geq - \frac{(2+z)^{2}}{4 L_{0}^{2}}\end{equation} Plugging in the value of $L_{0}$ from above gives\begin{equation}  m_{eff}^{2} \geq - \frac{3}{2 L^{2}}\,\frac{z+2}{z+1}\end{equation} Note that this expression interpolates between the bound for the case the IR geometry is  $AdS_{4}$ ($z=1$ and therefore $m_{eff}^{2} \geq - \frac{9}{4 L^{2}}$) and the bound for  the case the IR geometry is $AdS_{2}$ ($z=\infty$ and therefore $m_{eff}^{2} \geq - \frac{3}{2 L^{2}}$)\footnote{It may seem curious we reproduce the bound for the extremal charged black hole, which has $AdS_{2}$ in the interior but no scalar hair. This is the case because the BF bound depends only on the IR geometry, and not on how it is supported.}.

Another set of models we use was constructed by Horowitz and Roberts {\cite{Horowitz:2009p10812}}. In these models $m_{\psi}^{2}=0$, and for large enough charge, there is a superconducting order at zero temperature. The geometry in the IR is $AdS_{4}$ with the same radius of curvature as the UV space. The BF bound for a neutral scalar field $\xi$ is therefore identical in both limits of the geometry.

\subsection{Magnetic order in Reissner-Nordstr\"{o}m Backgrounds}

In \cite{Iqbal:2010p11098}, the authors discussed magnetic instabilities in the holographic context. In many models of condensed matter physics, the spin rotations decouple from spatial rotations at long distances, and can therefore be considered to be a separate global symmetry when discussing only IR physics. In the holographic context, such global symmetry is modeled by an SU(2) gauge field (which we denote as $A_{\mu}^{a}$). In order to model magnetic order, we are interested in configurations that break this global symmetry spontaneously. In particular, an anti-ferromagnetic order does not involve macroscopic background spin density, and can be modelled holographically by a scalar field in the adjoint $\psi^{a}$. By a suitable rotation, the symmetry breaking can be chosen to point in the 3 direction, and we denote the resulting neutral (with respect to the electromagnetic U(1)) scalar field $\psi^{3}=\xi$. The Action for the new fields $A_{\mu}^{a}$ and $\xi$ is \beq \mathcal{L} =  -\frac{1}{2\kappa^{2}}\left[G^{\mu\nu}G_{\mu\nu} +\frac{1}{\lambda}(\partial_{\mu}\xi\,\partial^{\mu}\xi +U(\xi))\right]\label{action}\eeq where $G_{\mu\nu}$ is the non-Abelian field strength, and $U(\xi)= \frac{1}{4}\left(\xi^{2}+m_{\xi}^{2}L^{2}\right)^{2} - \frac{m_{\xi}^{4}L^{4}}{4}$ is the potential for the scalar field $\xi$. To avoid clutter, we introduce the mass parameter $\nu= m_{\xi}^{2}L^{2}$.

The parameter $\lambda$ multiplying the action for the field $\xi$ controls the backreaction on the geometry. We will work in the probe limit, of $\lambda$ being infinite, thus avoiding such backreaction. The action for the excitations of the field $\xi$ will be bounded in all the backgrounds we consider, thus the probe approximation is justified (unlike an analogous approximation for the charged scalar fields).

In order to introduce finite temperature and density, the authors of \cite{Iqbal:2010p11098} work in a charged black hole background. This background is described by the metric: \beq
ds^{2}= L^{2} r^{2}\left( -f(r) dt^{2}+ dx^{2} +dy^{2} \right) + \frac{L^{2}}{r^{2}} \frac{dr^{2}}{f(r)}
\eeq where
\beq f(r) = 1 + \frac{3 \eta}{r^{4}}-\frac{1+3\eta}{r^{3}} \eeq and the U(1) gauge potential is $A_{0}= \mu (1- \frac{1}{r})$. The temperature and density are related to the parameter $\eta$ as
$\mu= \sqrt{3\eta}$ and $T= \frac{3}{4\pi}(1-\eta)$. The extremal limit corresponds to $\eta=1$.

The extremal Reissner-Nordstr\"{o}m background is a domain wall, interpolating between $AdS_{4}$ in the UV, and an $AdS_{2}$ space. Since the BF bound is different in the IR, there is a window of mass parameter, $-2.25<\nu<-1.5$ where the scalar $\xi$ induces instabilities in the IR, without destabilizing the complete geometry. In this range, we expect there to be a normalizable mode for the scalar $\xi$, signaling a symmetry breaking and an anti-ferromagnetic ordering. Such expectation is borne out by numerical construction of such normalizable mode, for mass parameters within the interesting range. The resulting phase diagram (as function of $\mu$ and temperature) is given in \cite{Iqbal:2010p11098}, and we reproduce it below in the appropriate limit of our discussion.

\section{Magnetic Instabilities in Holographic Superconductors}
\subsection{Phase Structure without Direct Coupling}
In this section we revisit the issue of magnetic instabilities, in the cases where the IR of the theory experiences condensation of charged scalar field, thus giving rise to superconductivity at low enough temperature (and in particular at T=0). We concentrate on the case where the IR theory developed dynamical scaling with scaling exponent $z$. The case $z=1$ gives rise to 2+1 dimensional conformal field theory in the IR. For the case $z=\infty$ has dynamical scaling in the time direction only, and the IR geometry is identical to the one discussed in \cite{Iqbal:2010p11098}.

\begin{figure}[h!]

\begin{center}
\includegraphics[scale=0.8]{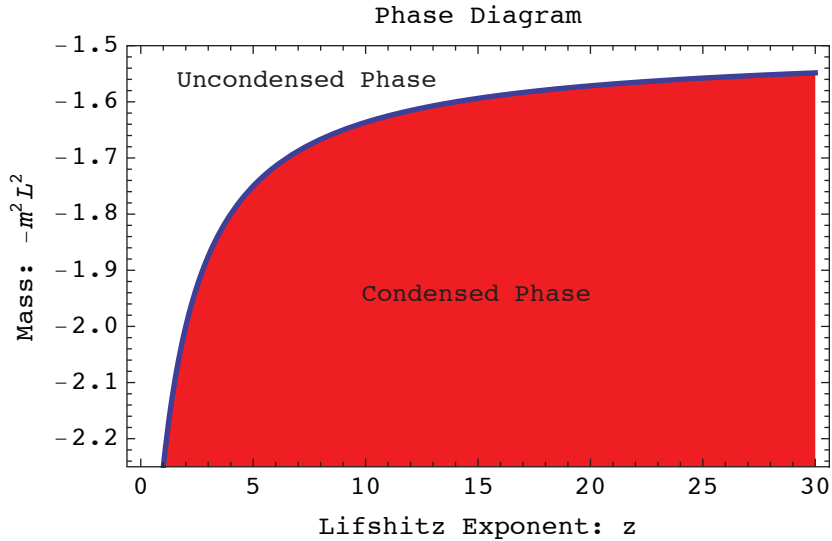}
\caption{Expected phase diagram, explicit solutions representing holographic superconductors were constructed numerically for low values of $z$. The expected phase diagram is confirmed for all those solutions.}
\label{BF}
\end{center}
\end{figure}

In figure (\ref{BF}) we draw the BF bound for the neutral scalar $\xi$ in the IR geometry. We see that the bound depends only on the dynamical exponent $z$. For any value of $z>1$, there is a range of masses for which the scalar $\xi$ is unstable in the IR region only, and as $z$ increases this window of instability increases.The expectation is therefore that magnetic order, in the sense defined in \cite{Iqbal:2010p11098}, is possible only in the region shown in figure  (\ref{BF}).

\begin{figure}[h!]

\begin{center}
\includegraphics[scale=0.8]{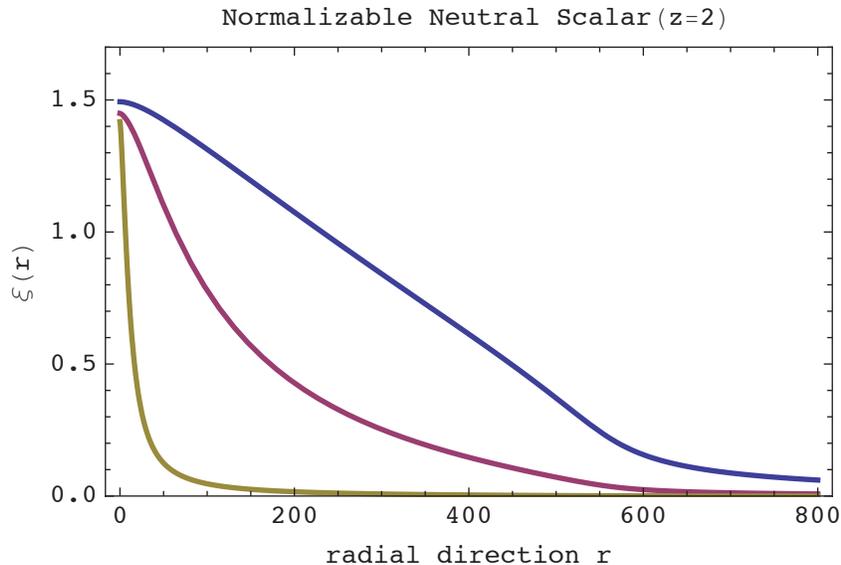}
\caption{Neutral scalar profiles for z=2 and various values of $\nu$ within the instability window $-2.25<\nu<-2$. The blue curve corresponds to $\nu=-2.23$, the red one for $\nu=-2.1$, and the yellow one for $\nu=-2.01$.
}
\label{norm}
\end{center}
\end{figure}

This expectation is borne out by explicit numerical calculations. First, the construction of Lifshitz solutions in {\cite{Gubser:2009p11021}} is an holographic superconductor only when embedded in an appropriate UV completion, representing a complete flow. We were able to construct such flows for low values of $z$ only. At asymptotically large $z$, the geometry becomes close to the extremal charged black hole background, which does not have scalar hair. We expect then that for sufficiently large $z$ there are no solutions which correspond to asymptotically $AdS_{4}$ flowing to the corresponding Lifshitz interior. In figure (\ref{norm}) we show the results obtained for the case $z=2$.

For that case (and several other low $z$ cases), we were able to construct normalizable solutions for the scalar $\xi$, if and only if the mass parameter $\nu$ is in the instability window.
For this range of masses, the fluctuations of the scalar $\xi$ around the maximum of the potential $U(\xi)$ becomes unstable in the IR region, and it is therefore close to the minimum of the potential, $\xi= \xi_{0}$.  On the other hand, the scalar
 is stable at the UV, therefore it is close to the maximum at $\xi=0$. The scalar profile is therefore an interpolation between these two values, facilitated by perturbing the IR value by an irrelevant perturbation, and the UV value by a relevant one. The ansatz for the scalar field is therefore \beq \xi(r) = \xi_{0} +a \,r^{\Delta_{{IR}}}\eeq for small $r$, and \beq \xi(r) = b \,r^{\Delta_{{UV}}}\eeq where $\Delta_{IR}>0$ is the IR dimension of $\xi$ around the minimum of the potential, and $\Delta_{UV}<0$ is the UV dimension of the field $\xi$ around the maximum of its potential \footnote{For the range of masses considered here, there is a choice of quantization in the UV, and our choice corresponds to the one considered in {\cite{Iqbal:2010p11098}}, it would be interesting to consider the alternate quantization, whose details a slightly more complex.}.

A few such scalar profiles are displayed in figure (\ref{norm}). For generic mass parameters within the instability window, the scalar profile lies entirely within the Lifshitz part of the geometry. Only when the mass parameter $\nu$ approaches the lower boundary of the instability window, $\nu = -2.25$, the profile becomes wider than the Lifshitz region, signaling instability of the entire spacetime. On the other boundary of the instability window, the solution becomes  infinitely thin, and thus within the errors of the numerical calculations.

The case $z=1$ corresponds to an emergent conformal symmetry in the IR. In such cases the instability window disappears, and we do not expect any normalizable solution. Indeed, numerical calculations fail to produce such solution, both for the set of background considered by Gubser and Nellore  {\cite{Gubser:2009p11021}}, and those considered by Horowits and Roberts {\cite{Horowitz:2009p10812}}.

The picture emerging from these numerical studies is that once the charged scalar condenses and superconductivity kicks in, the bulk geometry changes in such a way as to inhibit further instabilities, and in particular it inhibits the instability considered by {\cite{Iqbal:2010p11098}}.

\subsection{Adding Direct Couplings}
\begin{figure}[h!]

\begin{center}
\includegraphics[scale=0.8]{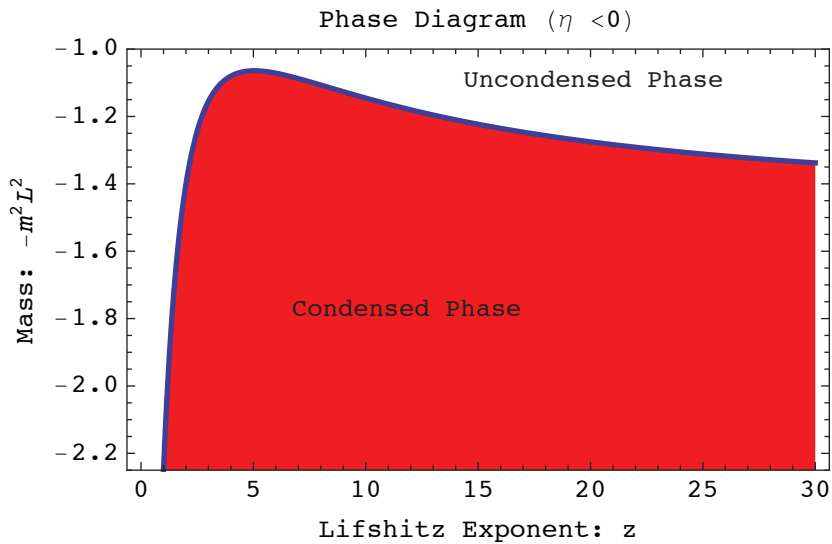}
\caption{Phase diagram with scalar interaction, and a negative coupling constant ($\eta=-0.3$).
}
\label{negative}
\end{center}
\end{figure}

While the results obtained  are fairly restricted, when the two scalars interact only gravitationally, much more general results can be obtained when the two bulk scalar fields interact directly.  Consider for example the modified action for the scalar field $\xi$ \beq \mathcal{L} =  -\frac{1}{2\kappa^{2}\lambda}\left[\partial_{\mu}\xi\,\partial^{\mu}\xi +\frac{1}{4}(\xi^{2}+\nu )^{2} - \frac{\nu^{2}}{4} +\eta |\psi|^{2} \, \xi^{2}\right]\eeq where $\eta$ is a new coupling which can be positive or negative. In the probe limit $\lambda \rightarrow\infty$ this coupling has vanishing effect on the charged scalar $\psi$ and the superconductivity it expresses. On the other hand, in the IR region (where $\psi=\psi_{0}$), this coupling translates into a shift of $2 \eta \psi_{0}^{2}$ in the effective mass of the neutral scalar field $\xi$. The condition for instability of that scalar in the IR becomes \beq
\nu< - \frac{3}{2 L^{2}}\,\frac{z+2}{z+1}-\frac{24 \eta}{m_{\psi}^{2}L^{2}} \, \frac{z-1}{(z+1)(z+2)}  \eeq
whereas in the UV $\psi=0$ and thus we still require $\nu> -2.25$. It is therefore clear that for $\eta<0$ we can obtain the situation where superconductivity increases the range of masses for which magnetic ordering occurs. Such situation is depicted in figure (\ref{negative}).

\begin{figure}[h!]
\begin{center}
\includegraphics[scale=0.8]{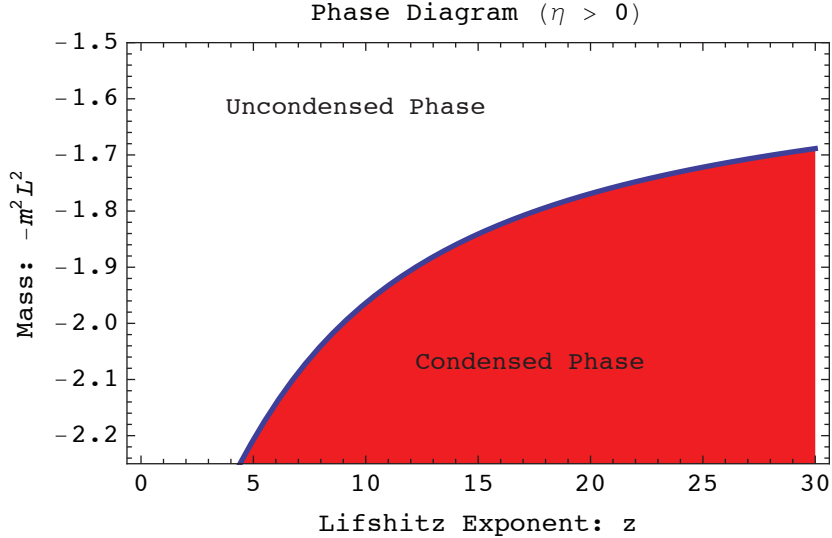}
\caption{Phase diagram with scalar interaction, and a positive coupling constant ($\eta=0.2$).
}
\label{positive}
\end{center}
\end{figure}

The situation is complementary for positive values of the coupling $\eta$. For general values of $z$ the coupling $\eta$ serves to further inhibit the magnetic ordering represented by $\xi$,  completely diminishing its window of instability for a range of scaling parameters $z$. In particular, at sufficiently low values of $z$ (which is the range more likely to correspond to complete  superconducting flows), the direct coupling dominates the effect due to the modified geometry, and prevent further instabilities for the scalar $\xi$. The phase diagram is shown in figure(\ref{positive}).

It is curious that in models with emergent CFT ($z=1$), such as the ones constructed by Horowitz and Roberts and used here, the addition of the quartic coupling discussed here, or any other similar couplings, has no effect. In such models superconductivity prevents any further ordering from occurring even with the inclusion of direct couplings between the order parameters.

The results in this subsection also clarify the situation occurring in the absence of any direct coupling. When the interactions between the holographic order parameters are attractive (as in the case of gravity, or a positive direct coupling), we find competition between the corresponding phases. Only when the interactions between the bulk fields are repulsive (such as the interactions mediated by vector fields), can both phases can coexist, and even enhance each other.

We conjecture that this correspondence holds in more general situations. Such correspondence could be useful as guide to modeling interesting phenomena for strongly coupled electrons, where cooperation between phases is expected to be an important part of the story.

\subsection{Superconductivity in Magnetically Ordered Phases}

We now turn to the complementary discussion, how is superconductivity affected once the holographic material enters the anti-ferromagnetic phase?

Suppose we include the backreaction of the scalar $\xi$ on the geometry, when the coupling $\lambda$ in (\ref{action}) is now finite. The detailed picture necessitates numerical solution of the coupled equations, but the leading order effect of the backreaction on potential superconductivity can be estimated by the following argument:

Since the scalar $\xi$ interpolates between its maximum $\xi=0$ in the UV, and it's minimum at $\xi=\xi_{0}$ in the IR, the leading order change in the geometry would be a shift of the cosmological constant in the IR, to be slightly larger than it was without the backreaction. The radius of curvature $L_{2}$ for the resulting $AdS_{2}$ would then decrease slightly.

Consider now the addition of a charged scalar field $\psi$ to the modified geometry, to model potential superconductivity.  A necessary  condition for instability of the $\psi=0$ configuration is \cite{Horowitz:2009p10812,Gubser:2009p11021}\beq -2.25<m_{\psi}^{2}L^{2}~~~~~\mbox{and}~~~~~(m_{\psi}^{2}- 2q^{2})\,L_{2}^{2}< -1.5 \eeq

The first condition is the BF bound in the UV, which is satisfied, and the second condition is the BF bound in the IR (using the effective IR mass $m_{\psi}^{2}- 2q^{2}$), which has to be violated for the $\psi=0$ configuration to be unstable, and thus for superconductivity to occur.

We find therefore that the leading order effect of the scalar $\xi$ is to modify the geometry is such a way as to make the second condition more restrictive, without modifying the first one. In other words, for any given charge $q$, the range of charged scalar masses which give rise to superconductivity shrinks due to the presence of the neutral scalar $\xi$.

This argument leads to results that are consistent with our general conjecture: for gravitational coupling only, or when including direct attractive interactions, the magnetic ordering seems to inhibit superconductivity. However, the addition of repulsive coupling may give rise to cooperation, and to superconductivity in the magnetically ordered phase, in cases where this would not occur without that magnetic ordering. It would be  interesting to look at this issue in more detail, either by constructing the back-reacted solution, or perhaps  by adding direct coupling between the two scalar fields. For example, models in which the anti-ferromagnetic phase enhances the possibility of superconductivity may be of relevance to the study of superconductivity in the cuprates.

\section{Global U(1) with two scalar fields}

In this section we generalize the discussion of holographic superconductors, to include two scalar fields coupled to a single Abelian gauge field\footnote{Models of quantum phase transition with two scalar fields were considered in \cite {Jensen:2010vx}.}. The two scalar fields are in competition, very much like the examples discussed above, and we will demonstrate, using both numerical and an analytic arguments, novel mechanisms by which a condensation of one scalar hinders the condensation of the other.

We therefore analyze the phase diagram of two scalar fields charged under the same $ U(1)$ gauge field. This corresponds to having two potential order parameters for a global symmetry in the boundary theory.  We will ignore their backreaction to the geometry since we are primarily interested in the onset of the condensations (i.e. we are working close to the point of the phase transition, where the amplitude of the condensate is small). The scalars will carry different charges and masses; since they do not back-react on the geometry, they are coupled only through the gauge field.

Consider then the planar limit of the four dimensional uncharged AdS black hole:
\begin{equation}
 ds^2 = -f(\tau) dt^2 + \frac{d\tau^2}{\tau^4 f(\tau)} +  \frac{1} {\tau^2}  (dx^2 + dy^2)
\end{equation}
with
\begin{equation}
f(\tau) =\frac{1}{L^2 \tau ^2}-M\tau.
\end{equation}
The horizon is  at $\tau=1$, while the conformal boundary lives at $\tau=0$, $L$ is the radius of the anti-de Sitter space and the temperature of the black hole is given by \footnote{It should be noted that "z" is used as a co-ordinate here. In the previous section it was used as Lifshitz exponent.}
\begin{equation}
 T =\frac{3M^{1/3}}{4\pi L^{4/3}}
\end{equation}
 In this note we will adopt the convention that $M =L=1$.

   The matter field Lagrangian is given by:
\begin{equation}
 L =\int dx^4 \sqrt{-g} ( -\frac{1}{4G}F^{a b}F_{ab}+ m_{1}^2\frac{|\psi_1 |^2}{L^ 2}+m_{2}^2 \frac{|\psi_2 |^2}{L^ 2} -|\partial\psi_1 -ie_1 A\psi_1 |^2-|\partial\psi_2 -ie_2 A\psi_2 |^2),
\end{equation} where $A$ is the $U(1)$ gauge field, and $\psi_{1},\psi_{2}$ are two charged scalar fields\footnote{From now on, any quantity defined with suffix 1 will be for scalar field $\psi_1$ and any quantity defined with suffix 2 would be for $\psi_2$. A quantity related to scalar field defined without a suffix will be a generic quantity relevant for both fields.}.

 We will also assume without loss of generality that $m_1^2 > m_2^2$.
We will be looking for static solutions and will assume all the fields are homogeneous in the field theory directions with only radial dependence.

The equations of motion for the fields in this coordinate system are:
\begin{eqnarray}\label{seom}
  \psi_1 ''&+&  \frac{ f'}{f} \psi_1 '+ \frac{1}{ \tau^4}   \left( \frac{e_1 ^2  A_t ^2  }{ f^2  }  -\frac{m_1 ^2}{  f} \right)\psi_1 =0\nonumber \\
\psi_2 ''&+&  \frac{ f'}{f} \psi_2 '+ \frac{1}{ \tau^4}   \left( \frac{e_2 ^2  A_t ^2  }{ f^2  } - \frac{m_2 ^2}{  f} \right)\psi_2 =0 \nonumber\\
  A_t ''&-&  2e_1 ^2\frac{ \psi_1 ^2} { f  \tau^4}   A_t'-  2e_2 ^2 \frac{ \psi_2 ^2} { f  \tau^4}   A_t =0.
\end{eqnarray}
By rescaling of $A_t$ we may always set the charge of one scalar to unity and in most cases we set $e_1=1$. Phase structure of the theory is determined by dimensionless ration $\frac{e1}{e2}$. 
To require regularity at the horizon we will have to set $A_t =0  $ at $\tau=1 $. Since we have a set of coupled equations, this will in turn give the constraints at the horizon 
\begin{eqnarray}
\label{regular}
  \psi_1 '&=&  \frac{-m_1 ^2 }{3} \psi_1\nonumber \\
  \psi_2 '&=&   \frac{-m_2 ^2 }{3} \psi_2  \nonumber\\
A_t&=&0
\end{eqnarray}
at $\tau=1$.
Examining the behavior of the fields near the boundary, we find
\begin{eqnarray}
  \psi_1 &\sim &  \Psi_{+,1} \tau^{\lambda_{+,1}}+ \Psi_{-,1}  \tau^{\lambda_{-,1}}  + ...\nonumber\\
   \psi_2 &\sim &  \Psi_{+,2} \tau^{\lambda_{+,2}}+ \Psi_{-,2}  \tau^{\lambda_{-,2}}  + ...\nonumber\\
 A_t &\sim &  \mu + \rho \tau^ {d - 2} + ...
\end{eqnarray}
The coefficients above can be related to physical quantities in the boundary  field theory using the usual dictionary in gauge/gravity correspondence. The constants $\mu $, $\rho$ are the chemical potential and the density of the charge carrier in the dual  field theory, respectively, and  $\lambda_{\pm,1,2} =\frac{1}{2}(d\pm \sqrt{d^2 +4m_{1,2}^2}) $.

As argued in \cite{Klebanov:1999tb}, for $m^2 \geq \frac{ -d ^ 2}{4}+ 1  $, only the term with $\lambda_+ $ is normalizable and only $\Psi_+ $ can be interpreted as the expectation value of an operator in a dual theory with dimension $\lambda_+ $. For $\frac{ -d ^ 2}{4}+ 1 \geq m^2 \geq \frac{ -d ^ 2}{4}  $, both terms are normalizable\footnote{Note that the conformal mass  is  the case $m^{2}= -2 $, for $d = 3$.}. However only one condensate can be turned on at a time in order to avoid instability in the asymptotic region of AdS. In this work we will confine ourselves to the case where only the mode with dimension $\lambda_+$ is tuned on.

\subsection{Qualitative Discussion}

The hair-less solution to equations eqn (\ref{seom}) is given by,
\begin{equation}
A_t=\mu \Big(1-\tau^{d-2}\Big), \ \  \psi_1=0, \ \ \psi_2=0
\label{eqn:nbkg}
\end{equation}

This is the normal phase solution that exists for all temperature. This solution is just the non-backreacted version of RN solution where back reaction of the gauge fields on the geometry is ignored.

Next, let us consider the case when value of one of the scalars is set to zero. The problem then reduced to the well-studied holographic superconductors \cite{Horowitz:2010p11086}. Let us assume that with the charge of the scalar fixed to unity, an instability in the scalar E.O.M. occurs at a critical value of $\mu=\mu_c$\footnote{In a conformal theory, only the ratio $\mu/T$ matters. Increasing the chemical potential is equivalent to lowering the temperature. In the following,  we choose to vary the chemical potential keeping the temperature fixed}. and a zero mode of the scalar field forms. If $\mu$ is increased further, the zero mode condenses and a new phase with a non-trivial profile for $\psi$ is formed. This is the well known mechanism for holographic superconductors \cite{Hartnoll:2008p1141,Gubser:2008px,Horowitz:2010p11086}.  $\mu_c$ is a increasing function of the mass square($m^2$) of the scalar.  As we vary the charge ($e$) of the scalars, $\mu_c$ is scaled to $\frac{\mu_c}{e}$. From the above discussion we may write down three solutions of the eqns (\ref{seom}),
\begin{itemize}
\item The hairless solution (eqn (\ref{eqn:nbkg})) exists for all value of $\mu$. And becomes locally unstable for $\mu>\mu_{c2}$.
\item The solution with $\psi_1=0$ and $\psi_2$ condenses. Exists for $\mu>\mu_{c2}$.
\item The solution with $\psi_2=0$ and $\psi_1$ condenses. Exists for $\mu>\mu_{c1}$.
\end{itemize}

Depending on the values of $e_2$ and $e_1$, the second or third solution may become locally unstable. There may be new phases where both of the scalars condense, or it may also happen that one of the solutions (either the second and third solution) simply dominates over the other.

To understand these issues in more detail we rewrite scalar EOM in (\ref{seom}) with a generic background $A_t$ as,
\begin{align}\label{psieom}
&\hspace{0.9in}\psi''+\frac{f'}{f}\psi'+\Big(\frac{e^2 A_t^2}{f^2}-{m^2 \over f})\psi=0 \\
\nonumber \Rightarrow &\hspace{0.2in}f(f \tilde \psi')'- V_{eff}(r)\tilde \psi =0,\quad \begin{cases}
\tilde \psi \tau^\frac{d-1}{2} = \psi \\
V_{eff}(r)=-f^2 \Big(-\frac{(d-1)(d-3)\tau^2}{4}+\frac{\tau^3 (d-1)f'}{2f}+\frac{e^2 A_t^2}{f^2}-{m^2 \over f}\Big)
\end{cases} \\
\nonumber \Rightarrow &\hspace{0.9in} \frac{d^2}{dy^2} \tilde \psi-\tilde V_{eff}(y)\tilde \psi = 0, \quad {\rm with} \quad dy=-\frac{d\tau}{\tau^2 f}.
\end{align}
Here $y\rightarrow\infty$ as $\tau \rightarrow 1$ and $y\rightarrow 0$ as $\tau \rightarrow 0$. In terms of this new variable $y$, the EOM of $\psi$ is rephrased as a potential problem on a semi infinite line ($y:[0,\infty)$). Depending on the nature of the potential $V_{eff}$,  there may exist a bound state of $\psi$. Such a bound state signifies an instability, suggesting that there may be new phases with non-trivial $\psi$. In the absence of  $A_t$, the scalar $\psi$ may condense if and only if $m^2 \leq m_{BF}^2$. This is the well known Bretinlohner-Freedman (BF) bound. Generically by choosing a sufficiently negative $V_{eff}$ one may force $\psi$ to condense, e.g. if we use the normal phase value of $A_t$ (eqn. \ref{eqn:nbkg}), and increase $\mu$ from zero, then eventually there will be zero mode of $\psi$. Increasing $\mu$ results in a bound state of $\psi$ signifying superconducting instability.

Due to the Higgs mechanism the condensation of one scalar depletes the gauge field from the IR region of AdS space. This in turn reduces the amount of negative potential contributing to the condensation of another scalar.  Hence we find that {\it generically condensation of one scalar hinders the condensation of another}\footnote{This was observed in the case of non-Abelian holographic superconductor in \cite{Basu:2008bh}.}. However whether or not a condensation of one scalar field will completely stop the condensation of the other is a question of detail.

From the above discussion it is natural to guess that, depending on whether  $\mu_{c2}/e_2 \gg \mu_{c1}/e_1$, the condensation of second scalar field dominates over the first one, or vice versa. Fortunately one may have some analytic arguments to understand these facts better and narrow down the range of possibilities.

\subsection{Analytic argument:}
Let us start by stating a simple lemma: if we define two potentials $V_1$ and $V_2$ over the same domain and $V_1 > V_2$ then the lowest eigenvalue of $V_1$ would be strictly greater than lowest eigenvalue of $V_2$. This simple fact may be proven using variational argument, as follows.

If the lowest eigenfunction in potential $V_1$ is $\psi_1$ with an eigenvalue $\lambda_1$,
\begin{align}
\lambda_1 &= \int  (-\psi_1 \psi_1''+ V_1 \psi_1^2) \nonumber \\
 &> \int  (-\psi_1 \psi_1''+ V_2 \psi_1^2)  \nonumber\\
 &\ge \int (-\psi_2 \psi_2''+ V_2 \psi_2^2) = \lambda_2
\end{align}

The above derivation implies if $V_2$ does not have a bound state solution, then $V_1$ cannot have one either. Also if the lowest eigenvalue mode for $V_2$ is a zero mode then $V_1$ can not have a bound state or a zero mode.

Being in $AdS$ complicates the picture but essence of the argument still survives. As we have discussed the scalar part of the Lagrangian may be written as a potential problem,
\begin{align}
S_{\rm scalar}&=-\int d\tau\ \frac{1}{\tau ^{d+1}}\left(m^2 \psi^2-\frac{1}{f}A_t^2\psi^2+f |\frac{1}{\tau ^2}\psi'|^2\right)  \nonumber \\
&=\int dy \tilde \psi (-\frac{d^2}{dy^2} \tilde \psi + \tilde V_{eff}(y)\tilde \psi)
\label{eqn:sbare}
\end{align}

Let us now get back to the two scalar problem. At first we will look at the case $e_2 > e_1$.

\subsubsection*{$\bf e_2>e_1$ case}
In this case we always have $V_{1eff}>V_{2eff}$\footnote{Remember $m_1^2>m_2^2$.}. No matter which gauge field configuration we choose, this fact remains true. The above argument implies then that a zero mode of $\psi_1$ may not form before a zero mode of $\psi_2$ does, as $\mu$ is increased from a small normal phase value. Consequently $\psi_2$ condenses before $\psi_1$. 

Now the question is whether a zero mode of $\psi_1$ may form in the phase where $\psi_2$ has already condensed. The answer is no, and the argument is as follows. It is to be noted that the $\psi_2$ condensate is a zero mode of $\psi_2$. This is just the retelling of the fact that a non-trivial node-less solution exists for E.O.M. of $\psi_2$.  Existence of such a solution comes primarily from numerics. The condensation of $\psi_2$ changes the gauge field profile and correspondingly change the potential problem. In this modified potential problem the condensate  is a zero mode of $\psi_2$.  This enables us to use the lemma proven at the beginning of this section. Since with the modified (i.e. modified by the condensation of $\psi_2$)  gauge field  $V_{1eff}>V_{2eff}$\ and $\psi_2$ has a zero mode, it follows from the lemma that no zero mode or bound state of $\psi_1$ exists. Hence in the phase with a $\psi_2$ condensation, $\psi_{}1$ can not condense. This is in accordance to our general expectation: once $\psi_{2}$ condensed, it depletes the gauge potential, thus changing the effective potential for $\psi_{1}$ from one that allows a bound state, to one that does not.

The slight modification of the above argument actually prevents any state with simultaneous condensation of $\psi_1$ and $\psi_2$, as both may not have zero modes in any gauge field configuration. Moreover any phase where $\psi_1$ has a condensed will necessarily have a instability for $\psi_2$. Hence for $e_2>e_1$, any phase with non-zero $\psi_1$ is dominant. The phase structure of the system is therefore the same as that of a single scalar holographic superconductor with the scalar field $\psi_2$.

Next, let us consider the converse situation, where $ e_2 \le e_1$.
\subsubsection*{$\bf e_2 \le e_1$ case}
The situation is more complicated here. Looking at the Lagrangian, one may naively expect that as we increase the chemical potential the scalar with more charge will always eventually dominate. However the situation is more complicated as the potential $V_{eff}$ in eqn (\ref{psieom}) diverges like $1 \over y^2$ near the boundary $y=0$. In this case the above argument is essentially reversed and the mass dependent potential part possibly becomes important as we turn on more chemical potential. This concurs with the numerical finding discussed next.

\subsection{Numerical Analysis}

Here we study the eqn (\ref{seom}) numerically. We confine ourselves to the case with $m_1^2=0$ and $m_2^2=m^2_{conformal}=-2$ (see \cite{Horowitz:2008bn}). With just one of the scalar turned on and the charge of it set to unity we know that the scalar condenses as the value of $\mu$ is increased from zero. Let us assume that without any interface from another scalar, scalars condense at $\mu=\mu_c$. If we set the charge of the scalars to unity then $\mu_{c1} \approx=7.8 \approx 1.9 \mu_{c2} (\approx 4.0) $.  As the charge ($e$) of the scalar varies from unity, $\mu_c$ is scaled to $\mu_c/e$.

As discussed before in the regime $e_2 > e_1$ the phase diagram is same as that of a model with only second scalar. The same picture holds in the regime $e_1 < \frac{\mu_{c1}}{\mu_{c2}} e_2 \approx 1.9 e_2$ and the second scalar dominates the picture and there is no condensation of the first scalar.

\begin{figure}\label{fig:twoscalar}
\begin{center}
\includegraphics[scale=0.8]{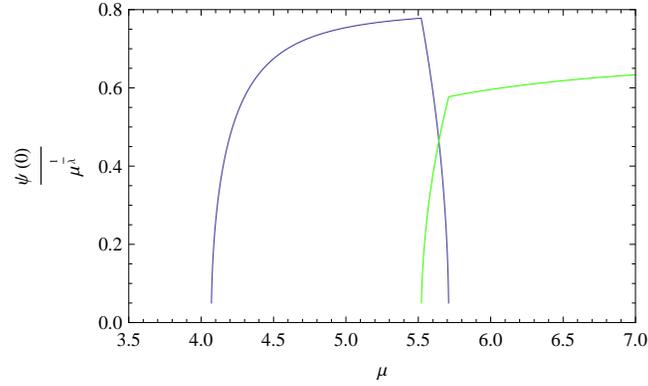}
\end{center}
\caption{Plot of condensate with two scalars with $e_1=1.95$ and $e_2=1$. The curve on the right is for the second scalar and it starts from $\mu=\tilde \mu_{c2}$. The left one is first scalar it starts from $\mu=\mu_1$ and ends at $\mu=\tilde \mu_{c1}$. In the range $\tilde \mu_{c2}<\mu<\tilde \mu_{c1}$ both of the scalar condenses. }
\end{figure}

As $e_2$ decreases beyond $\frac{\mu_{c2}}{\mu_{c1}} e_1$ the first scalar condenses before the second scalar. As we increase the boundary value of chemical potential more the second scalar condenses at a chemical potential $\mu=\tilde \mu_{c2}$. The condensation of second scalar increases with chemical potential. As argued in the previous section increasing chemical potential possibly makes the mass dependent potential term more important and hence the second scalar tends to dominate the picture. As $\mu$ is increased further the condensation of first scalar decreases and goes to zero at $\mu=\tilde \mu_{c1}$. Although the first scalar condenses before the second one, one may consider the solution where the expectation value of first scalar is kept to zero and let the second scalar condenses. As chemical potential is increased, there will be some value $\tilde \mu_{c1}$ when the first scalar will develop a zero mode. Numerically $\tilde \mu_{c2}$ is close to $\tilde \mu_{c1}$ and consequently one phase quickly changes to another. In this case there are three apparently second order transition at the increasing order of $\mu$ (see fig. \ref{fig:twoscalar}):
\begin{enumerate}
\item A second order phase transition at $\mu=\mu_{c1}$ when first scalar condenses.
\item A second order phase transition at $\mu=\tilde \mu_{c2}$ when second scalar start condensing. The resulting phase has both scalar condensates.
\item A when the condensation of first scalar goes to zero at $\mu=\tilde \mu_{c1}$.
\end{enumerate}
Numerically it seems that as $e_2$ is decreased even further such that $e_2 < 1.15 \frac{\mu_{c1}}{\mu_{c2}} e_1 \approx 2.2 e_1$ the second scalar never condenses and the resulting phase diagram is same as that of model with only first scalar. Whether this is a numerical artefact or not needs to be investigated in more detail.

In principle there may exist first order transitions in our system. The particular case we have discussed seems to have none. A detailed study with other possible values of mass and dimension is left out for a future work.

\subsection{Effect of Scalar Interactions}
Similar to Lifshitz case discussed above, turning on a direct interaction between scalars should help scalars to condense simultaneously, in case that interaction is repulsive. It is then expected that the regime of co-existence of phases will be enhanced. We expect an opposite effect for a repulsive interaction.  Detailed study of these issues are left for a future work.

\section*{Acknowledgements}

We thank Josh Davis, Kristan Jensen, Tommy Levi, Hong Liu, Shamit Kachru  and Mark van Raamsdonk for useful conversations and correspondence. The work is supported by discovery grant from NSERC.
\bigskip
\bigskip
\bigskip
\bigskip

\newpage
\bibliography{refs}

\end{document}